# *Small* Large-Scale Wireless Networks: Mobility-Assisted Resource Discovery

Ahmed Helmy
Department of Electrical Engineering
University of Southern California
helmy@usc.edu

*Abstract* - In this study, the concept of small worlds is investigated in the context of large-scale wireless ad hoc and sensor networks. Wireless networks are spatial graphs that are usually much more clustered than random networks and have much higher path length characteristics. We observe that by adding only few random links, path length of wireless networks can be reduced drastically without affecting clustering.

What is even more interesting is that such links need not be formed randomly but may be confined to a limited number of hops between the connected nodes. This has an important practical implication, as now we can introduce a distributed algorithm in large-scale wireless networks, based on what we call *contacts*, to improve the performance of resource discovery in such networks, without resorting to global flooding.

We propose new *contact*-based protocols for adding logical short cuts in wireless networks efficiently. The new protocols take advantage of mobility in order to increase reachability of the search. We study the performance of our proposed contact-based architecture, and clarify the context in which large-scale wireless networks can be turned into *small* world networks.

*Index terms* - Small world graphs, resource discovery, ad hoc and sensor networks, wireless networks, mobility analysis.

## I. INTRODUCTION

The concept of small worlds was studied in the 60's in the context of social networks[1], during which experiments of mail delivery using acquaintances resulted in an average of 'six degrees of separation', i.e., on average a letter needed six acquaintances to be delivered. Recent research by Watts[2][3] has shown that in relational graphs adding a few number of random links to regular graphs results in graphs with low average path length and high clustering. Such graphs are called small world graphs.

Emerging multi-hop wireless networks, such as ad hoc and sensor networks, do not belong in the category of relational graphs. Rather, they belong in the category of spatial graphs, where the links between nodes depend on the radio range, which in turn is a function of the distance between the nodes. The applicability of small world graphs to spatial graphs has not been established by [2].

In this paper, we attempt to develop a better understanding of the small world concept in the context of wireless networks. Specifically, what are the path length and clustering characteristics of wireless networks? Can we create networks that have reduced path length by adding a small number of short cuts? Do these short cuts have to be totally random? What do these short cuts represent given the limited radio connectivity in wireless networks? It would likely represent logical connections that translate into multiple physical hops. Finally, given reasonable answers to the above questions, can we develop network architectures, for resource discovery, that incur low degrees of separation between nodes in large-scale wireless networks?

Our study attempts to provide answers to the above questions. Under the assumptions of our study, results in this paper suggest that we can actually reduce the path length of wireless networks drastically by adding a few random links (resembling a small world). Furthermore, these random links need not be totally random, but in fact may be confined to a small fraction of the network diameter, thus reducing the overhead of creating such network.

Based on these observations a new architecture is introduced that attempts to create a small world in large-scale wireless networks (with potentially thousands of nodes). The architecture is based on defining *contacts* for network nodes. These contacts are chosen in a simple and efficient way by taking advantage of mobility. The goal of such contacts is to be used during resource discovery without global flooding.

We analyze the different parameters of our architecture and evaluate its performance in terms of network reachability. We also specify a new set of mobility-based protocols for efficient contact selection. We evaluate our proposed protocols under mobility conditions. Initial results indicate promise for efficient contact selection, and we propose several favorable protocol settings based on our analysis. In future work, we plan to conduct more detailed evaluations using different mobility models and richer metrics. We shall also investigate other heuristics for contact selection.

The rest of the paper is outlined as follows. Section II provides background on small world concept and outlines our work in that area. Section III presents experiments and results for small world wireless networks. Section IV provides a link between our small world analysis and resource discovery in wireless networks. Section V presents related work in the area of resource discovery. Section VI introduces our contact-based architecture. Section VII introduces and analyzes a new class of mobility-based contact selection protocols. Section VIII concludes and discusses future work.

## II. SMALL WORLDS

The *small world* phenomenon comes from the observation that individuals are often linked by a short chain of acquaintants. Milgram [1] conducted a series of experiments to study social contacts and networks. In his experiments, individuals were randomly chosen to deliver letters between Nebraska and Massachusetts, the source can only deliver the letter through acquaintants knowing the target's address and occupation. It was found that an average of 'six degrees of separation' exists between senders and receivers. This provided an interesting yet intriguing result. What is the structure of social networks that allows to create a small number of degrees of separation within a very large population? In an effort to further understand such structure Watts [2] conducted a set of re-wiring experiments on graphs ranging from regular graphs to random graphs. It was observed that by re-wiring a few random links in regular graphs, the average path length was reduced drastically (approaching that of random graphs), while the clustering[1] remains almost constant (similar to that of regular graphs). This class of graphs was termed *small world graphs*, and it emphasizes the importance of the random links acting as *short cuts* that contract the average path length of the graph. The experiment was conducted for relational graphs, in which links are not restricted by distance between the nodes. It was also noted that for spatial graphs, in which links are a function of the distance between nodes, the small world phenomenon does not exist (that is, path length and clustering curves almost match)[2]. According to the experiment, in such graphs it is hard to add short cuts while maintaining clustering.

Multi-hop wireless networks, including ad hoc and sensor networks belong to the class of spatial graphs, where the links are determined by the radio connectivity, which is a function of distance. Hence, we expect that such networks, by their own nature, do not lend themselves to small worlds. We also expect high clustering in wireless network due to the locality of the links, since many of a node's neighbors may be neighbors of each other too. Also, due to this locality we expect the path length for such networks to be high as compared to randomly connected networks.

In this work, we conduct further experiments on spatial graphs in the context of multi-hop wireless networks, and investigate the applicability of the small world concept to these networks. Our study takes a practical perspective in which we hope to utilize small worlds in designing efficient protocols for ad hoc and sensor networks. In particular, we propose a novel contact-based architecture for resource discovery in large-scale ad hoc and sensor network. In such architecture, our goal is to reduce the number of queries during the search for a target node or resource. The distance, in terms of number of queries, between the source of the search and the target is called degrees of separation. Based on the notion of *contacts*, we create logical short cuts to reduce the degrees of separation. We study the number of needed short cuts and how they should be distributed across the network.

## III. SMALL WORLDS SIMULATION AND ANALYSIS

We start our experiments by investigating several layouts of wireless networks. Without loss of generality, we choose a setting of 1000 nodes over a 1kmx1km area. We investigate various distributions of node placement, including random, normal, skewed and grid. Some of the topologies are shown in Figure 1. Several values of radio ranges were chosen to provide different number of links and average node degrees. A subset of the topologies used in our study is given in Table 1.

Re-wiring experiments were conducted on the above networks, where a node is chosen at random and one of its neighbors is chosen at random, then the link between them is removed and re-linked to a random node. Because this may result in network disconnection, which may affect our results, we also performed link

---

[1] By *clustering* we refer to the *clustering coefficient* that basically captures the fraction of nodes' neighbors that are also neighbors of each other. Clustering refers in some sense to the underlying structure of the network.

[2] Some of Watts' experiments were conducted on 1-dimentional spatial graphs, in which $k$ links, on average, originate from every node. Links for all nodes were chosen within a distance $d$. Path length and clustering exhibited similar dynamics as $d$ increased.

addition experiments, where 2 nodes are chosen at random and a link is added between them.

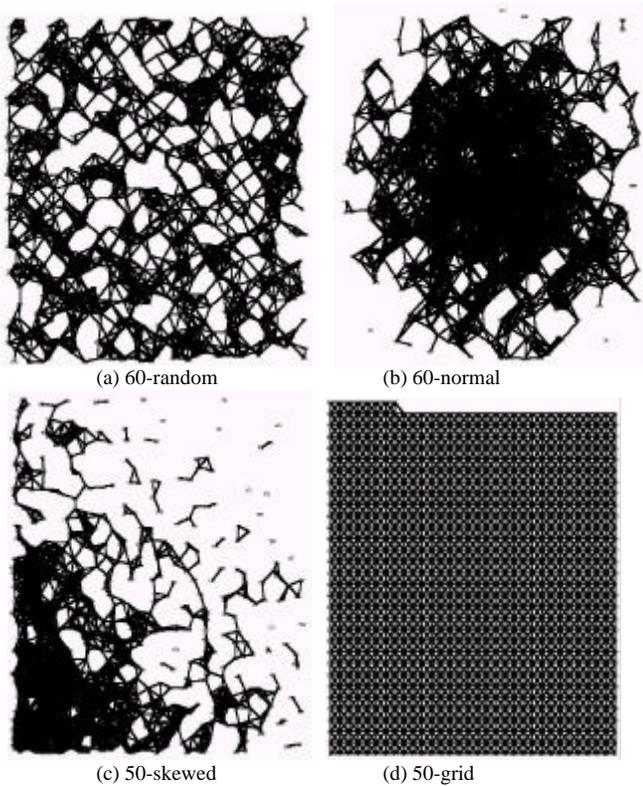

**Figure 1** Some topologies used in the study: (a) randomly located nodes with 65m range, (b) nodes located with a normal distribution (avg. 500, std. 250) with 65m range (c) nodes located with a skewed distribution with 55m range, (d) nodes located on a grid with 50m range.

| Topology | Range (m) | Links | Topology | Range(m) | Links |
|---|---|---|---|---|---|
| 55-random | 55 | 4785 | 35-grid | 35 | 1936 |
| 65-random | 65 | 6850 | 50-grid | 50 | 3811 |
| 65-normal | 65 | 10790 | 75-grid | 75 | 9310 |
| 55-skewed | 55 | 10051 | 100-grid | 100 | 12872 |

**Table 1** Topologies used in our study

Re-wiring and link addition experiments were conducted for various numbers of links (or probability of re-wiring/addition). For every probability of re-wiring or link addition, $p$, the average path length, $L$, maximum path length, $m$, and clustering coefficient, $C$, are measured. For the original case, $p=0$, (without re-wiring or link addition) these values are denoted as $L(0)$, $m(0)$ and $C(0)$, respectively. For other values of $p$ we get $L(p)$, $m(p)$ and $C(p)$, respectively. We plot the ratios $L(p)/L(0)$, $m(p)/m(0)$, and $C(p)/C(0)$ on a semi-log plot. These ratios represent reduction in length or clustering with increased probability of re-wiring or link addition. Some of the results are shown in Figure 2. For all the other experiments we got strikingly similar results.

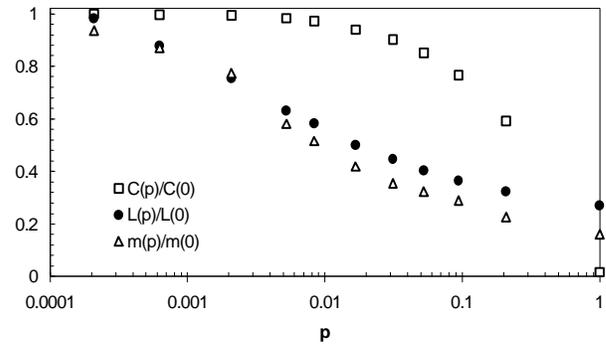

Figure 2. Reduction of path length and clustering vs. probability of re-linking

We note several observations on these results. First, the values for clustering and path length of the original graphs (before re-wiring or adding links) are quite high as compared to those of random graphs. These values are shown in Table 2. One exception is for 35-grid, where $C(0)=0$. Aside from this exception, wireless networks, in general, tend to be highly clustered (as we expected) due to the locality of the links, which increases the probability that a node's neighbors are also neighbors of each other.

|  | Rand graph | 55-rand | Normal | Skewed | 35-grid | 50-grid |
|---|---|---|---|---|---|---|
| $C(0)$ | 0.009 | 0.58 | 0.568 | 0.567 | 0 | 0.45 |
| $L(0)$ | 3.3 | 12.3 | 6.9 | 8.92 | 21.1 | 14.8 |
| $m(0)$ | 5 | 31 | 21 | 32 | 62 | 31 |

**Table 2** Clustering, path length and maximum length for the original networks without re-wiring or link addition ($D$ is the network diameter)

Second, from the Figure, we observe a very consistent trend among all the experiments and across all topologies. There is a clear distinction between the reaction of the path length and clustering to re-wiring or link addition. The path length reduction occurs quite drastically for 0.2% to 20% of re-wiring. Further re-wiring does not contribute much to reducing the path length. We found that on average, re-wiring or addition of 0.2% of the links results in 25% reduction in $L$, re-wiring/addition of 2.58% results in 50% reduction and re-wiring/addition of 19.8% results in 70% reduction.

Finally, there is a very clear difference between the path length and clustering dynamics with link re-wiring or addition. For example, to achieve 75% for $L(p)/L(0)$ we need around 0.2% re-wiring, but to achieve 75% $C(p)/C(0)$ we need around 9% re-wiring, i.e., there are two orders of magnitude difference for the needed re-wiring. This suggests that by adding or re-wiring a very small number of links we can drastically reduce the path length while not affecting the structure of the underlying network. These results seem consistent with the small world graph phenomenon.

If we consider reduction in clustering as a measure for the change in the structure of the network, then it would be interesting to investigate the point of maximum reduction in path length reduction with minimum re-structuring of the network. To achieve this, we define the metric *C/L* equal to *C(p)/C(0)/(L(p)/L(0))*, and the metric *C/L/(C/L)max* to normalize *C/L*. In Figure 1, we plot this value for the various graphs under study and the relational graphs (that were studied in[2]). It is quite interesting to see the trend of such value with the probability of re-wiring or link addition, *p*, as shown in the Figure. Although the values may not be the same for all *p*, the trend is very similar, with the maximum point around *p*=3% to 9%. This shows that, up to a point, significant decrease in path length may be achieved while maintaining the clustering structure of the network. Beyond such point minor reduction in path length requires significant change in the clustering and hence the structure of the network.

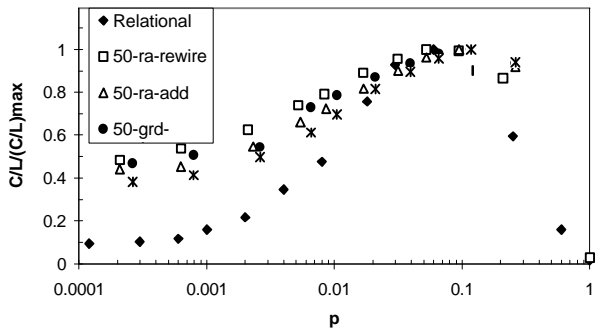

Figure 3 Normalized *C/L* ratio with max around 3%-9%

The previous analysis shows that, for the studied networks, it seems quite possible to achieve significant reduction in degrees of separation in wireless networks by adding a few (say 0.2%-2%) *random* links. What has not been clear this far is how random should the *random* links be? From a practical point of view, choosing random contacts for a node in an ad hoc or sensor network may result in unpredictable overhead to select and maintain routing for that contact. So the next step would be to investigate the possibility of limiting the distance from which a contact may be chosen (such that the overhead is more predictable), while still achieving good path length contraction. The next set of experiments address exactly this issue.

For this set of experiments we add a small number of links that achieved a large decrease in path length. We choose three values for the added links: 25, 80, and 150 links (these values achieved between 40-60% reduction in *L* in our previous experiments). We want to obtain the link (or contact) distance that achieves the largest reduction in *L*. In these experiments, we limit the maximum distance (in hops) from which a contact maybe chosen by a node, call this distance *r*. A contact may be chosen at random from a distance *d*, where 2<*d*<*r*. We vary *r* from 2 to the network diameter *D*. Results are shown in Figure 4. The experiments show a clear and consistent trend for all topologies[3]. The path length reduction, *L(p)/L(0)* seems to increase with the increase of the distance *r* up to a certain fraction of network diameter *D*, then it saturates. Further increase in the distance *r* does not seem to add much to the path length contraction. We measured the fraction *r/D* after which further increase did not affect the reduction by more than 3%. The average of such fraction for the topologies studied was around 45% (with minimum of 35% and maximum of 50%).

Since the contact is chosen randomly from [2,*r*] hops, we expect the average contact distance, $d_{av}$, to be around 20-35% of the network diameter in order to achieve the most significant and effective reduction in path length. Note that the probability distribution is actually weighted by the number of nodes at each hop, this turns out to be a bit higher than 2 + [r-2]/2. This result is further discussed later.

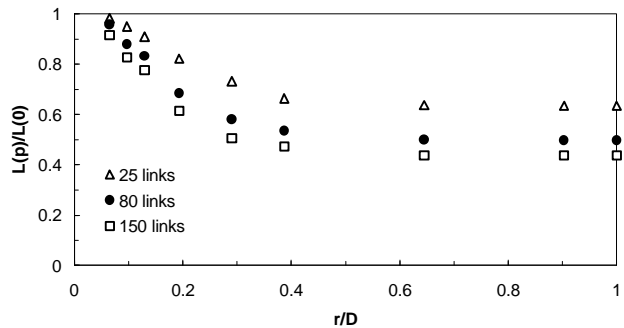

Figure 4 Path length reduction vs. max contact distance *r* normalized by network diameter *D*.

---

[3] The Figure shows results for 55-random. Very similar trends were observed in the other topologies.

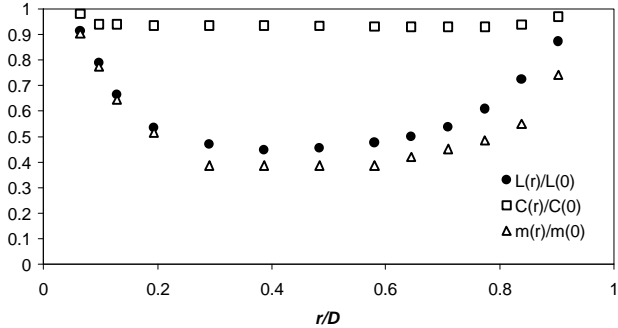

Figure 5 Choosing contacts from exactly *r* hops. Decrease in path length is at a maximum around 25-40% of the network diameter *D*.

The next set of experiments involved choosing the contacts from exactly *r* hops away and investigating the trends in path length and clustering. We wanted to see if we will get consistent results with the previous experiments. The results are shown in Figure 5 (similar results obtained for all topologies). To our surprise, the curve does *not* saturate after a certain distance. What is observed is that at a certain distance (roughly between 25-40% of the network diameter) the maximum reduction in path length is achieved, after which the path length *increases*. Perhaps this occurs according to the node distribution with hops where the contact reaches the maximum number of nodes around that certain distance, on average. Nonetheless, this seems to be a result with very significant practical impact. This indicates very clearly that by limiting the number of hops between the added links (or contacts) to just a fraction of the network diameter (say 25%) we can achieve maximum reduction in path length or degrees of separation. This result, along with the previous observations encouraged us to pursue our contact-based architecture described in the rest of this document.

*Discussion*

Previous work done by Kleinberg [4] suggests that, in a 2-dimensional grid, the best message delivery time measured in number of forwarding steps (which corresponds to degrees of separation in our case) is achieved when the short cuts are chosen with a probability inversely proportional to their distance from a node. For example, for two nodes *A* and *B* separated by distance $d(A,B)$ hops, the probability of *B* being a contact for *A* is $d(A,B)^{-r}$, where *r* is variable. More specifically, the best performance occurs when $r=2$, as the expected number of forwarding steps is at most proportional to $(\log n)^2$ (with each node knowing about only its neighbors and one contact). Note that in order to obtain this probability in an ad hoc network each node needs global knowledge of the locations of all other nodes in the network (or at least prior knowledge of the node distribution with hops w.r.t. each node). This is not practical. However, if this probability is per hop, a node may identify itself as *d* hops away from another node (using TTL or similar), and can compute this probability independently with a rough estimate of the network diameter, *D*, in a decentralized way.

So, what we want to study is the distance (in hops) that achieves the best performance. Note that $d^2$ used in[4] is per node (not per hop). We want to obtain the expected number of hops for such probability distribution. Given the 2-dimentional grid and given a general node near the center of the grid, the number of nodes that can be reached at exactly *d* hops away is 4*d* (approximate for the regular grid and exact for the grid on a sphere). Taking the probability of each node to be $1/d^2$ and that 4*d* nodes are at hop *d* we get the hop probability distribution as *1/d* (i.e., the probability of choosing the contact at hop *h* is 1/*h*). To get the probability distribution we normalize by Σ*1/d*, for all hops up to the network diameter. Hence the expected hop value is $D/(\Sigma 1/d)=D/H(D)$, where *D* is the network diameter, and Σ*1/d* is the harmonic series, *H(D)*. $H(D)=ln(D)+O(1)$. Around 30-70 hops the ratio ranges from around 20-25% of *D*, consistent with our results above. We also conducted simulations to get the per-hop distribution and obtained the expected number of hops for the various topologies under study with an overall average of 25.5% of *D* (min 21%, max 34%) [very much in lines with our results above.].

Next, we attempt to make use of these results to design a novel contact-based architecture for resource discovery in large scale ad hoc and sensor networks.

IV. SMALL WORLDS & RESOURCE DISCOVERY IN LARGE-SCALE WIRELESS NETWORKS

Now that the relationship between small worlds and wireless networks has been established, we need to ask the question 'How do we make use of such information to design better architectures and protocols for ad hoc and sensor networks?' First, now we have a good estimate of the number of short cuts to be established in order to achieve significant path length reduction. We also have a good estimate of the distance of these links. So, how can we establish those links in wireless networks? and what do they mean? We shall address the latter question first. Short cut links in relational graphs may be obtained practically by adding a physical (wired or other) link between the chosen nodes. In ad hoc networks, however, the physical constraint posed by the transmission range renders the ad hoc network a 'spatial graph', which cannot include random links naturally. Note that short cuts may be achieved by increasing the transmission range to establish the link. This however, may also be limited in distance, causes increased power consumption, and has negative effects on utilization of the spectrum. Some systems may exploit several radios, one with short range (and high data rate) and another with longer range (and low data rate), which may establish the short cut. This may be a possible system to investigate in the future. In this study, however, we do not assume that such capability exists and hence our short cut (for the remainder of this document) will be a logical link that will translate into several physical hops. The logical link, in our case, corresponds to one degree of logical separation.

Thus, our goal is *not* to reduce the physical path length. Rather, we aim to reduce the logical path length (in terms of degrees of separation) and in turn reduce the number of queries during resource discovery. This, along with the insights gained from our experiments (such as the number and placement of contacts) establishes a clear link between the previous part of the paper on small worlds and the coming part on resource discovery in ad hoc and sensor networks.

Also, note that the above analysis, for $L, C$ and $m$, showed the existence (and context) of the small world characteristic, using short cuts, in wireless networks. It did *not* show, however, 'how to find and use those short cuts'. This question is only answered through architectural and protocol design; a subject we shall discuss for the remainder of this paper.

## V. RELATED WORK ON RESOURCE DISCOVERY IN WIRELESS NETWORKS

We address the problem of resource discovery in *infrastructure-less* networks. The resource discovered may even be a route. Hence, architectures that require infrastructure (e.g., DNS) or that assume existence of underlying routing are not suitable for our problem. Centralized approaches are neither robust nor scalable.

In wireless ad hoc and sensor networks perhaps the simplest form of resource discovery is global *flooding*. This scheme does not scale well. It also uses broadcast, which is usually unreliable at the data link layer (e.g., in 802.11). The synchronization of the broadcasts may lead to severe collisions and medium contention. Hence, it is our design goal to avoid global flooding. Expanding ring search uses repeated flooding with incremental TTL. This approach also does not scale well. Much of the work on routing protocols in ad hoc networks uses some form of flooding or ring search (e.g., DSR[7], AODV[6], DSDV[5], ODMRP[8]).

Other approaches in ad hoc networks that address scalability employ hierarchical schemes based on clusters or landmarks (e.g., LANMAR[10] and [9]). These architectures, however, require complex coordination between nodes, and are susceptible to major re-configuration (e.g., adoption, re-election schemes) due to mobility or failure of the cluster-head or landmark. Furthermore, usually the cluster-head becomes a bottleneck. Hence, in general we avoid the use of complex coordination schemes for hierarchy formation, and we avoid using cluster-heads.

In GLS [11] an architecture is presented that is based on a *grid* map of the network (that all nodes know of). Nodes recruit location servers to maintain their location. Nodes update their location using an ID-based algorithm. Nodes looking for location of a specific ID use the same algorithm to reach a location server with updated information. This is a useful architecture when a node knows the network grid map, knows its own location (through GPS or other), and knows the ID of the node it wishes to contact. These assumptions may not hold in our case. Especially that a source node has to know the specific ID of the target node and uses that ID to locate the traget's location servers. By contrast, in our architecture, a source node may be looking for a target *resource* residing at a node with an ID *unknown* to the source node.

The algorithm proposed in [4] and [12] uses global information about node locations to establish short cuts or friends, and uses geographic routing to reach the destination. It is unclear how such architecture is feasible with mobility. Also, such work does not specify the number of short cuts to create. In addition, the destination ID (and location) must be known in advance, which may not be the case in resource discovery.

In ZRP [13] the concept of hybrid routing is used, where table-driven routing is used intra-zone and on-demand routing is used inter-zone. Border-casting (flooding between borders) is used to discover inter-zone routes, which may not scale well. A good feature in ZRP is that a zone is node-specific. Hence, there is no complex coordination susceptible to mobility as in cluster-head approaches. We use the concept of zone in our architecture. However, we avoid border-casting by using contacts out-of-zone.

In [14] an architecture based on intelligent agents is introduced for resource discovery in ad hoc networks.

The concept of domains is used and a cluster-head election is needed to define a domain.

## VI. CONTACT-BASED ARCHITECTURE

First, we state the assumptions upon which our architecture is built.

*Assumptions:* (1) The source does not know the ID of the target node that holds the resource. (2) Nodes only have local knowledge of their neighbors (e.g., using 1 hop Hello or data link connectivity). (3) Nodes do not know their own location or any other geographical location of any other node (i.e., our architecture does not require GPS or any other GPS-less distance estimation techniques)[4]. (4) Infrastructure-less network: We assume there are *no* well-known servers or landmarks. (5) We assume rough knowledge of network size (number of nodes) at the design stage.

These assumptions differentiate our work from other works that require any of the above elements.

*Architecture:* We propose what we call a loosely coupled simple hierarchy. It is loosely coupled because each node has its own view of the network (its zone), and it is simple because there are no complex coordination mechanisms to elect cluster-heads or leaders. In our architecture, each node knows a number of neighboring nodes in a neighborhood or zone. The zone may be defined, for example, by a number of hops $R$. Routes to nodes within the zone are established by a local table-driven intra-zone routing protocol (e.g., DSDV). Outside of the zone, a node may maintain (a small number of) contacts. A contact is chosen at $r$ hops distance from the selecting node. The contact may be maintained up to $r_{max}$ hops away. We presented a very high level sketch of this idea in [15] without details. The idea behind the contacts stems from the small world concept, but also for increased coverage and reachability of other nodes. A contact outside a node's zone will also have its own zone, thus providing an extended view of the network.

The idea behind our architecture is simple. However, several challenging questions impose themselves. What should be the value of the zone radius, $R$? What should be the contact distances $r$ and $r_{max}$? How many contacts should be established? How will the query mechanism be implemented on top of such architecture? And, perhaps one of the most important issues, how will these contacts be chosen and how will mobility affect this architecture? We leave this latter question for detailed treatment in the next section. To answer the first few questions about $R$, $r$, contact number (*CN*) and the query, we perform an evaluation of the design space. Also, we define new parameters for the query, the query depth, *QD*, and the contact level. For a node, its contacts are level-1 contacts. Then its contacts' contacts are level-2 contacts, so on. The contact depth is the number of levels of contacts that the query is allowed to reach, in order to search for the target. For example, a scheme in which a node queries only its contacts for the target is a *QD=1* scheme.

For brevity, we show our analysis of the 55-Rand topology shown earlier. Similar analysis may be performed for other topologies as well.

Simulations were run for 1250 queries (from 25 random sources, each to 50 random targets), and were averaged for different random seeds. Figure 6, shows the un-reachability (i.e., the %ge of failed searches for a target), as function of $R$ and the query depth. Our findings show that the parameter that affects reachability the most is the query depth. This is because the number of contacts searched grows exponentially with *QD*. Also, $R$ has significant effect on reachability. As $R$ is increased, the number of nodes reached in-zone grows significantly. For $r$ it turns out that so long as $r$ is around $2R$ or greater, reachability does not change drastically. The reason for this is at $r > 2R$ we get no overlap between the source's zone and the contact's zone.

We also need to consider overhead of changing $R$, *QD* and $r$. The greater these values, the greater the overhead of intra-zone routing, querying and contact maintenance, respectively. So, we choose min values that achieve very good reachability. From our results we pick *R=4, QD=4, r=7* (this also coincides with our small world finding as 22% of $D$ (31)).

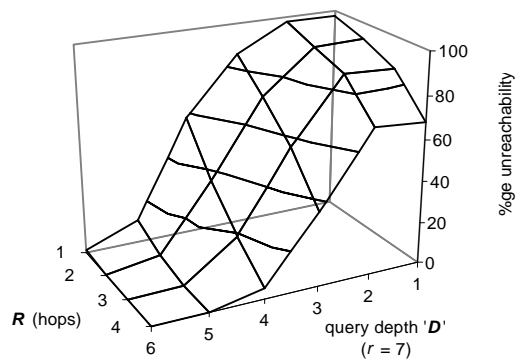

Figure 6 Reachability as function of zone radius and query depth.

---

[4] Availability of such location information may simplify our architecture. Studying such simplifications is part of on-going and future work.

For the number of contacts *CN* per node (not shown for brevity), the effect on reachability is nearly linear. However, we do not recommend to increase the number of contacts per node. Rather, we should introduce the notion of contact sharing among nearby nodes. This reduces the overhead of contact selection and maintenance. Issues of contact sharing are part of ongoing and future work. For now, we may intuitively recommend between 80-150 contacts for the whole topology (these were numbers that achieved very good path length reduction in our small world analysis). We shall also investigate a lazy (or on-demand) approach for contact selection in the future. We shall also research the generality of our findings to other topology configurations (with varying number of nodes and transmission ranges), and investigate *scaling factors* that easily translate these results from one topology to another. Another future issue is dynamically configuring these parameters. In this study we address parameters setting at the design stage, assuming that the rough network size is known. In the next section we address the issue of contact selection.

VII. MOBILITY-BASED CONTACT SELECTION PROTOCOLS

One of the main challenges that we need to address in our architecture is the effect of mobility and its dynamics on contacts. We pose this challenge in the form of the following question. How will contacts be selected with reasonable overhead, while achieving the small world characteristics under mobility conditions? To attempt to answer this question we first address the question of contact selection and present various protocols that address mobility issues. Then we evaluate the effect of mobility on selected contacts. Our goal in this study is to gain better understanding of the interplay between contacts and mobility.

*A. Contact Selection*

As was pointed out earlier, according to our analysis a *desirable region* for a contact to be in (for our studied topologies and those similar) is around 20%-25% of the network diameter, *D*. How can a node select a contact so many hops away with reasonable overhead (without resorting to global flooding)? At the same time, we want the selection process to elect contacts that are also likely to remain in the *desirable region* long enough to be useful in resource discovery. How can the selecting node choose the contacts intelligently, given that the desirable region will be out-of-zone. This problem is challenging for two main reasons. First, mobility seems as an adversary, providing random movement and contributing to link and path failure. Second, the selecting node is likely to know little or no information about the mobility characteristics and capabilities of nodes in the desirable region.

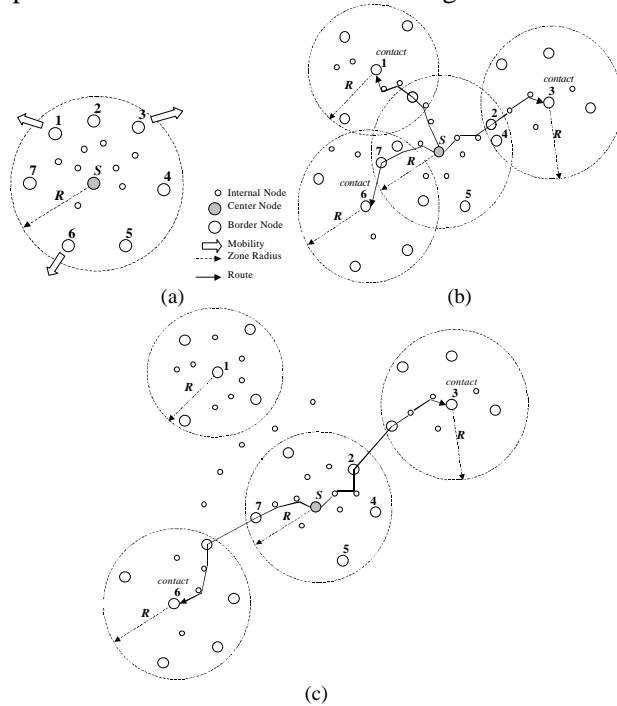

Figure 7 Example of zoning, contacts and effect of mobility: (a) Zone for source node *S* is shown (with radius *R*). Border nodes are numbered (1-7). Nodes 1,3 and 6 are moving/drifting out of zone. (b) Radii for the drifting nodes are shown. *S* stays in *contact* with the drifting nodes, which enables it to obtain better network coverage with low overhead. (c) After moving away, contact nodes drift up to a point where their zones no longer intersect with *S*'s zone. In this example, *S* maintains contact with those nodes not more than (2*R*+1) hops away, i.e. nodes 3 and 6, and loses contact with node 1 as it drifts farther than the contact zone.

Our proposed answer to the above question is to *take advantage of mobility* using a novel approach called mobility-based contact selection. In our approach, a selecting node, *S,* makes initial selection of a list of *candidate contacts (CCs)* from its own zone. These are nodes that lie within ***R*** hops away. *S* knows routes to these nodes via intra-zone routing. This way, *S* may also collect information about *CCs'* mobility or abilities. This information may be piggybacked over the intra-zone route exchange. The future *contacts* are to be chosen from this list of *CCs*. Once these candidate contacts move out-of-zone, overlap between their zones and *S's* zone will be reduced and the added network view (or coverage) will be significant, as shown in Figure 7. Still, some questions remain regarding the dynamics of these contacts. Will they move into the desirable region and for how long? How long will they remain candidates before they do enter the desirable region? To get a better understanding of the effect of mobility on our proposed approach, we define two different protocols for mobility based contact selection.

Let us start by defining the overall scheme of contact selection, then move into the details of the individual protocols. In both protocols, a node, *S,* starts selecting a list of candidate contacts *(CCs)* from the zone (i.e., within *R* hops). Based on its mobility pattern, a node on the *CCs* list may get promoted to *contact* or get dropped (or evicted) from the candidacy list based on *promotion/eviction* rules. By mobility pattern we mean the sequence of distances (in hops) between these candidates and the selecting node, over time. The protocols differ in the way they select candidates from the zone, and the way they use the mobility pattern in conjunction with the promotion/eviction rules.

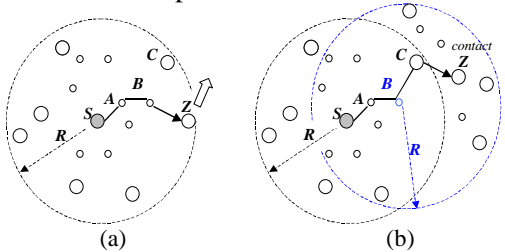

Figure 8 The zone information propagation (*ZIP*) local route discovery protocol (a) *Z* is at the border of *S*'s zone with a route '*S-A-B-Z*'. *Z* is moving out-of-zone. (b) *Z* is no longer within *S*'s zone, but it highly likely to exist in *B*'s zone. The route '*S-A-B-C-Z*' is identified and *Z* is selected as a *contact.*

As these *CCs* move, *S* keeps track of their distance (in hops) through intra-zone routing, and when they move out-of-zone a lightweight local route discovery mechanism, called zone information propagation (*ZIP*) is used (see Figure 8)*.* When and if a candidate crosses the *promotion boundary (PB)* it is considered a *contact* and may be queried during searches. Once a contact crosses the upper or lower *eviction boundaries (UEB, LEB)* it is evicted from the *contact list* and is not queried in further searches (see Figure 9). That is, if a node moves too far away (beyond the eviction boundary) it is harder to maintain, whereas if it comes closer to *S* its new zone may overlap with that of *S,* and is evicted in both cases.

The two selection protocols are called *border-based and neighbor prediction-based contact selection protocols*.

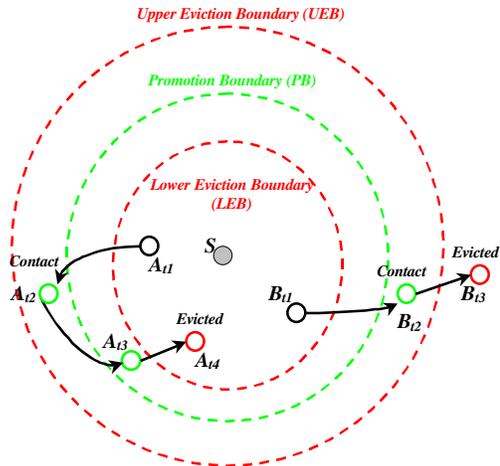

Figure 9 Promotion and Eviction boundaries for *S:* nodes *A* and *B* originally in *S*'s zone, get promoted at time t2 when they cross the promotion boundary (PB), then they get evicted from the contact list at time t3 when they cross the lower or upper eviction boundary.

*Border based contact selection protocol:* Selection of candidate contacts is done from nodes at *R* hops, i.e., at the border of the zone. Nodes *R* hops away are tracked, i.e., *S* keeps track of its movement/distance over time. If border nodes move closer to *S*, they are evicted. If they cross the promotion boundary, those candidates become contacts. When the contacts cross an eviction boundary they are evicted.

*Neighbor prediction based contact selection protocol:* This protocol takes advantage of the fact that *S* readily knows the routes/hop distance to nodes within its zone (i.e., within *R* hops away). It also takes advantage of the likelihood that (even in random way point movement) mobility often remains constant for short whiles. In this protocol, node *S* selects neighbors (those nodes that are 1 hop away) to *track* their movement. When a neighbor node becomes at 2 hops away, then 3 hops away, this sequence may indicate that this neighbor is heading out of zone and has a high probability to continue moving until it becomes in the desirable region (the probability of this happening is a function of the mobility pattern and is part of this study). Other neighbors, that do not show this consistency in mobility, may take longer (if ever) to get to the desirable region and thus waste more resources before becoming useful, and so are evicted from the *CCs* list. We expect this simple prediction scheme to help identify good candidates that have higher probability of becoming contacts. We plan to investigate such behavior in our simulations and analysis.

## B. Evaluation under mobility

First, we define our evaluation metrics based on our study questions. We define the *contact persistence*, as the life span of a contact (the point in time when the candidate contact became a contact until the time it was evicted). In general, the more the persistence the better, since we want the contact to remain in the desirable region for a longer time and be useful for more queries. The second metric we define is the *average overlap range*. For each source node choosing contacts, this metric measures, at each instance in time, the pair-wise zone overlap distance of its contacts, normalized by the number of pairs (say if we have $n$ contacts, then $n(n-1)/2$ is the normalization factor). This is averaged for all the sources in the simulation run. It may also be averaged over the simulated time to get the average of the averages. This metric attempts to capture the coverage overlap between the contacts. The less the better. Another metric is the expected time for a node to become a contact (the faster the better). Yet others include the number of contacts (or %ge of *CCs* list becoming contacts), number of contacts evicted by the upper boundary vs. lower boundary (the more at the upper boundary the better, than means they went through the desirable region), among others.

For mobility we use the random way point model, in which a node chooses a random destination and picks a random velocity from [0,Vmax]. Once the destination is reached another destination and velocity are picked, so on. We use a pause time of 0, i.e., continuous mobility.

We investigate several dimensions in the design parameter space. For both protocols, we vary the promotion and eviction boundaries systematically and observe the change in the above metrics. We vary the number of neighbors/borders chosen initially by each protocol. We also vary the max velocity used. For brevity we only show results for one topology, 1000 node, 1kmx1km, 55-Rand described earlier (similar analysis maybe carried out similarly for other topologies). We use $R=4$. Note that $r$ now is a function of the promotion and eviction boundaries. We vary the promotion boundary (PB) and the lower and upper eviction boundaries (LEB, UEB), sometimes allowing hysteresis (i.e., PB > LEB). PB and LEB are varied between 4 and 7 and UEB is varied between 10 and 13 (this is the space of parameters we were most interested in analyzing based on our previous analyses).

The simulations were run for 100 and 200 seconds, with Vmax ranging from 1m/s to 40m/s, using 50 randomly selected sources (*S*), tracking 7, 14, 20 borders each (for the first protocol) and all $1^{st}$ hop neighbors (for the second protocol). We discuss a subset of our results showing the most interesting trends.

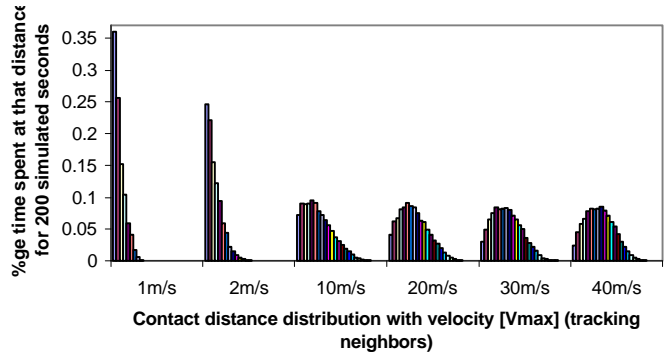

(a) tracking neighbors

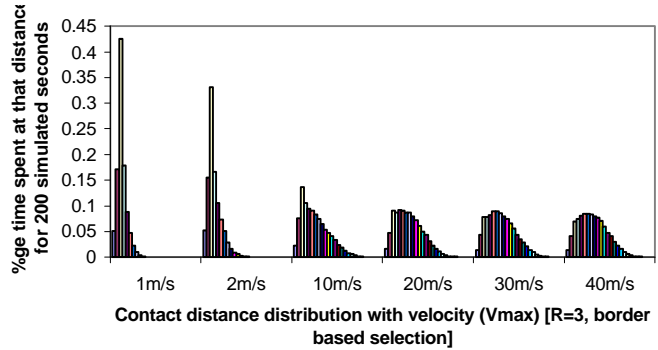

(b) tracking borders

Figure 10 Tracking experiments

To get a feel of the node movement, especially the relative distance (in hops) between a selecting source node (*S*) and a tracked node (candidate contact or contact), we present a set of *tracking* experiments without boundaries. Since the two protocols track different nodes (the first tracks border nodes and the second tracks $1^{st}$ hop neighbors) we conduct both experiments. As shown in Figure 10 (a) and (b), the eventual distribution of relative distances seems to be quite similar for both protocols and it saturate around 20m/s and above. The graphs show that, in general, the nodes spend much of their time between 4 and 9 hops away from their corresponding *S*. The distribution is very similar to that of the node-spread distribution (or node-hop distribution) for that topology. This seems promising because of the reasonable probability that a tracked node will spend time in the desirable region. Also, we notice that, over time, for the tracked borders much time is spent in hops closer to *S* than the border (i.e., hops 1, 2). This may indicate a notable probability of border nodes moving closer to *S* before moving out-of-zone. We shall investigate this effect further as part of the experiments below. The graphs do not show, however, the movement pattern for individual nodes. Such pattern is captured in the next set of experiments.

We now study the effect of varying the promotion boundary (PB) and the lower eviction boundary (LEB). As shown in Figure 11, (for the neighbor prediction protocol and 20m/s Vmax), the number of contacts selected increases with the decrease in PB. Also, the number of remaining contacts decreases with the increase in LEB. Contact persistence drops from an average of 42.9sec (PB=4,LEB=4) to 27.15sec (PB=7,LEB=4) to 18.5sec (PB=7,LEB=7). On the other hand, the average overlap range was as follows: 23.2m (PB=4,LEB=4), 15.9m (PB=7,LEB=4), 12.5m (PB=7,LEB=7). We notice that overlap increases with number of contacts and with lowering PB. With decreased PB the contacts will be relatively closer to *S* and hence have higher probability of being close to each other as well. With increase in LEB we expected to see a significant drop in overlap. However, the drop was minor. This indicates that most contacts crossing PB moved to farther relative distances. To follow up on this observation we look at the number of contacts evicted at LEB vs. those evicted at UEB. Recall that being evicted at UEB is preferable because the contact will have crossed the whole desirable region. The numbers were as follows (for the same overall number of selected contacts, 301 contacts): LEB=4 achieved 50 LEB evictions and 235 UEB evictions (16 contacts remained at the end of simulation), whereas LEB=7 achieved 83 LEB evictions and 210 UEB evictions (8 contacts remained). Hence all the extra evictions due to lowering LEB seem to have been for useful contacts (that would have otherwise crossed over to UEB and been UEB evictions, or remained as contacts). Hence, allowing the contacts (elected at 7 hops) to remain contacts between 4-7 hops, gives them a highly probable opportunity to move on to the desirable 7-10 hops region without crossing the 4 hops boundary. This argues in favor of the 'hysteresis' scheme (PB=7,UEB=4). From the above analysis we recommend a setting of (PB=7,UEB=4).

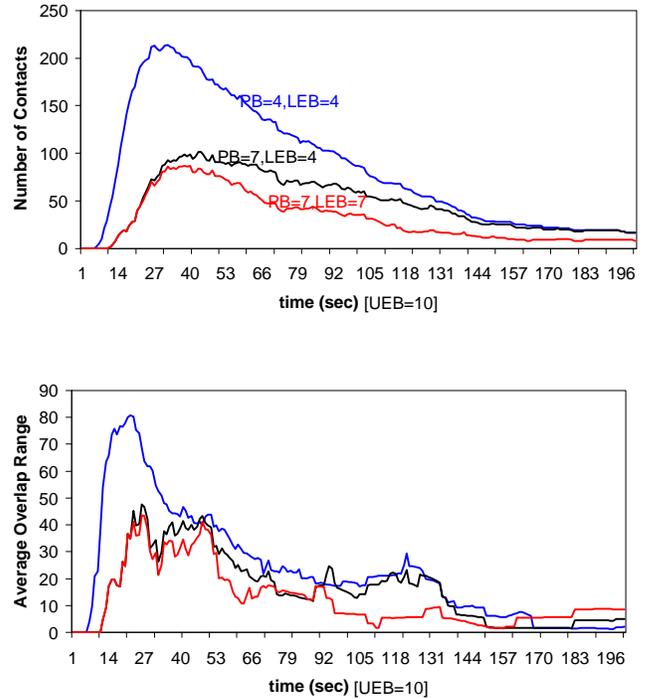

Figure 11 Effect of varying the promotion boundary (PB) and the lower eviction boundary (LEB) for the neighbor prediction protocl.

Next we study the effect of changing the upper eviction boundary (UEB). As shown in Figure 12, the number of remaining contacts increases with the increase in UEB, and so does the average number of contacts over time. In addition, contact persistence increases from 27.15sec (UEB=10) to 36.5 (UEB=11) to 42.19 sec (UEB=12) to 49.53 (UEB=13). We also notice an increase in average overlap with increase in UEB, from 15.9m (UEB=10) to 18.9 (UEB=11) to 21.6 (UEB=12) to 23.97 (UEB=13). (Remember that the zone radius is *R*=4 and the transmission range is 55m). This increase is probably partially due to increase in average number of contacts. It may also be due to the fact that after a specific number of hops, contacts tend to cluster near the boundary of the network, which reduces their relative distances. The increase does not seem significant though. Hence, it seems from this analysis that the higher UEB the better. What is missing from this data, however, is the change in maintenance overhead with increased UEB. This needs more detailed description of the protocol and is part of our future work.

Next, we compare the performance of the two different contact selection protocols. We discuss our results for PB=7, LEB=4, UEB=10, Vmax=20m/s, and 300 selected contacts[5]. For the boundary-based selection we notice that the expected time for contact promotion (10.25sec) is much lower than that in the neighbor prediction scheme (45.18 sec). We also notice that a much larger %ge of tracked nodes that converted into contacts was much higher in the neighbor prediction scheme (80%) than in the border selection scheme (54%). This is due to the filtering scheme that occurs during prediction, where nodes not conforming to the prediction rule are not tracked at the boundary of the zone. Average overlap was very similar in both cases. Contact persistence, however, was much higher in border selection (45.8sec) than in neighbor prediction (27.15sec). This could be due to the faster promotion in the former scheme. Hence, if the border tracking can be achieved without more overhead than neighbor tracking then it seems that border selection scheme is better. Tracking overhead analysis is part of our future work.

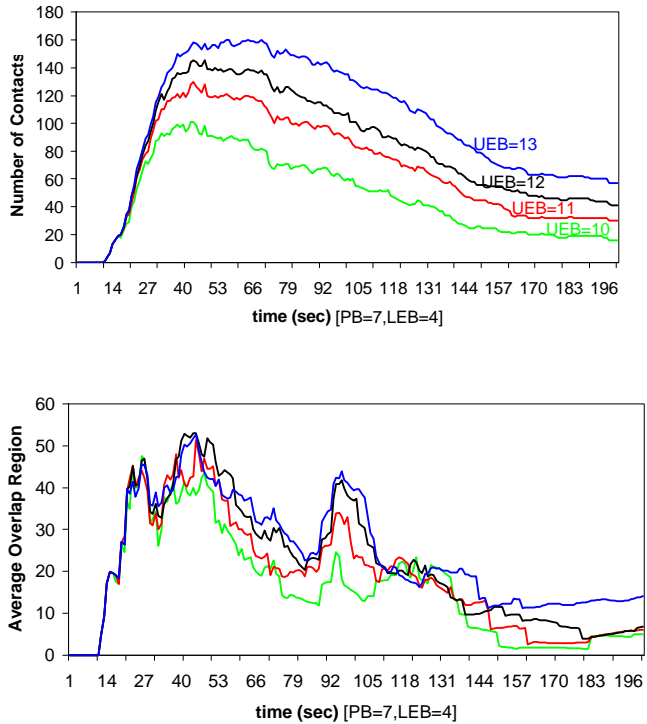

Figure 12 Effects of varying the eviction upper boundary (UEB)

VIII. CONCLUSIONS AND FUTURE WORK

We have established a relationship between small world graphs and wireless ad hoc and sensor networks. Our findings indicate that by adding a few short cuts, with only a small fraction (~20%) of the network diameter, the degrees of separation may be reduced drastically. This was established both through extensive simulations and through analysis. Based on this finding we proposed a novel contact-based architecture for resource discovery in ad hoc and sensor networks. To realize such architecture efficiently, we introduced a new class of protocols called mobility-based contact selection protocols, including two selection protocols, local route discovery protocol and a set of promotion/eviction rules. Our experiments on contact selection protocols and node tracking enabled us to develop a base-line understanding of the mobility dynamics and its effect on our architecture. Initial results indicate promise of the proposed schemes, and point to favorable parameter settings for our protocols. Future study is needed to further enhance our understanding and investigate more traditional metrics for evaluation (such as coverage and overhead). We plan more detailed evaluation of our schemes using various mobility models (e.g., group mobility, freeway models, etc.). We also plan to investigate other heuristics for contact selection based on minimum overlap rules.

REFERRNCES

[1] S. Milgram, "The small world problem", *Psychology Today* 1, 61 (1967)
[2] D. J. Watts. In *Small Worlds, The dynamics of networks between order and randomness*. Princeton University Press, 1999.
[3] D. Watts, S. Strogatz, "Collective dynamics of 'small-world' networks", *Nature* 393, 440 (1998).
[4] J. Kleinberg, "Navigating in a small world", *Nature,* 406, Aug. 2000.
[5] C. E. Perkins, P. Bhagwat, Highly Dynamic Destination-Sequenced Distance Vector Routing (DSDV) for Mobile Computers, Comp. Commun. Rev., Oct. 1994, pp. 234-44.
[6] C. E. Perkins, E. M. Royer, Ad-hoc On-Demand Distance Vector Routing, Proc. 2$^{nd}$ IEEE Wksp. Mobile Comp. Sys. And Apps., Feb. 1999, pp. 90-100.
[7] D. B. Johnson, D. A. Maltz, Dynamic Source Routing in Ad-Hoc Wireless Networks, Mobile Computing, 1996, pp.153-181.
[8] S. Lee, M. Gerla, C. Chiang, "On-demand multicast routing protocol", IEEE WCNC, p. 1298-1302, vol. 3, 1999.
[9] B. Das, V. Bharaghavan, "Routing in ad-hoc networks using minimum connected dominating sets", *in Proc. IEEE ICC '97.*
[10] P. Guangyu, M. Gerla, X. Hong, "LANMAR: landmark routing for large scale wireless ad hoc networks with group mobility", MobiHOC '00, p. 11-18, 2000.
[11] J. Li, J. Jannotti, D. Couto, D. Karger, R. Morris, "A Scalable Location Service for Geographic Ad Hoc Routing", ACM Mobicom 2000.
[12] L. Blazevic, S. Giordano, J.-Y. Le Boudec "Anchored Path Discovery in Terminode Routing". *Proceedings of the Second IFIP-TC6 Networking Conference (Networking 2002), Pisa, May 2002.*
[13] M. Pearlman, Z. Haas, "Determining the optimal configuration for the zone routing protocol", IEEE JSAC, p. 1395-1414, vol. 17, 8, Aug 1999.
[14] J. Liu, Q. Zhang, W. Zhu, J. Zhang, B. Li, "A Novel Framework for QoS-Aware Resource Discovery in Mobile Ad Hoc Networks", *IEEE ICC '02*.
[15] A. Helmy, "Architectural Framework for Large-Scale Multicast in Mobile Ad Hoc Networks", *IEEE ICC '02*.

---

[5] This provides a fairer comparison since the number of selected contacts depends on the number of candidates (for first protocol may be much more than 1$^{st}$ hop neighbors).